# Automated Volumetric Segmentation of Microaneurysms on OCT Using Artificial Intelligence

Min Gao[1,2], Yukun Guo[1,2], Tristan T. Hormel[1], Jinyi Hao[1], Azaz Khan[1,2], Steven T. Bailey[1], Thomas S. Hwang[1], Yali Jia[1,2, *]

[1] Casey Eye Institute, Oregon Health & Science University, Portland, OR 97239, USA

[2] Department of Biomedical Engineering, Oregon Health & Science University, Portland, OR, USA

## Abstract

**Purpose:** To develop and validate a deep learning-based method for the automated identification and volumetric segmentation of microaneurysms (MAs) in diabetic retinopathy (DR) using optical coherence tomography (OCT).

**Design:** Cross-sectional study with a representative longitudinal case analysis.

**Participants:** A total of 125 participants were enrolled, including 20 healthy eyes, 27 with mild nonproliferative DR (NPDR), 30 with moderate NPDR, 30 with severe NPDR, and 18 with proliferative DR.

**Methods:** We obtained multiple repeated 3×3-mm scans from each participant using a commercial 120-kHz spectral-domain OCT system (Solix; Visionix/Optovue, Inc., California, USA) which were registered and averaged to generate high-definition volumes. We developed an end-to-end three-dimensional convolutional neural network, incorporating an encoder-decoder architecture with residual blocks, to segment MAs volumetrically. The input to the network consists of the original OCT volume concatenated with its reflectance-inverted counterpart to enhance feature extraction. Expert graders manually delineated MAs to generate initial annotations, which were later refined into a verified ground truth through an iterative human-in-the-loop refinement process to capture MAs initially overlooked or mischaracterized. Both single-scan volumes and volumes averaged across 4-10 repeated OCT volumes were used to train and test the model. We evaluated model performance at multiple levels, including voxel-level segmentation, lesion-level detection, and eye-level diagnosis. We also analyzed the differences in MA counts across various stages of DR severity and quantified the MA number and volume distribution in different retinal layers. A longitudinal analysis of MA dynamics was performed in a case with severe NPDR and macular edema using the algorithm.

**Main Outcome Measures:** Voxel-level MA segmentation accuracy, MA quantification categorized by retinopathy severity, and longitudinal analysis of MA dynamics in OCT volumes.

**Results:** In the test dataset (20 healthy, 20 DR eyes), when evaluated against the verified ground truth, the algorithm demonstrated high voxel-level accuracy, with F1 scores of 79.2% on single volumes and 86.1% on averaged volumes. Lesion-level detection reached an F1 score of 96.0% (single) and 97.1% (averaged), while scan-level diagnostic accuracy for MA presence was 97.5% (single and averaged). MA counts significantly increased from mild to moderate NPDR (7.4±1.5 vs. 12.7±1.6, P = 0.014) but not between any other severity comparisons (P>0.05). MAs were primarily localized to the inner nuclear layer and outer plexiform layer. In one case with severe NPDR and macular edema, longitudinal analysis tracked 11 resolved MAs and 3 new MAs post-treatment, corresponding to a reduction in fluid volume.

**Conclusions:** A deep learning-based method can accurately identify and segment MAs volumetrically on OCT, enabling quantification and characterization and potentially aiding in the diagnosis, monitoring, and management of DR.

**Funding:** This work was supported by grants from National Institutes of Health (R01 EY 036429, R01 EY035410, R01 EY024544, R01 EY027833, R01 EY031394, R43EY036781, P30 EY010572, T32 EY023211, UL1TR002369); the Jennie P. Weeks Endowed Fund; the Malcolm M. Marquis, MD Endowed Fund for Innovation; Unrestricted Departmental Funding Grant and Dr. H. James and Carole Free Catalyst Award from Research to Prevent Blindness (New York, NY); Edward N. & Della L. Thome Memorial Foundation Award, and the Bright Focus Foundation (G2020168, M20230081).

**Financial interests:** Optovue/Visionix (P, R), Genentech/Roche (P, R, F), Ifocus Imaging (I, P), Optos (P), Boeringer Ingelheim (P, R, F, C). Yukun Guo: Visonix/Optovue, Inc. (P), Genentech, Inc. (P, R). Tristan T. Hormel: Ifocus Imaging (I).

## 1. Introduction

Microaneurysms (MAs) are often the first clinically detectable sign of diabetic retinopathy (DR), appearing before more advanced vascular abnormalities. Their presence is a key criterion in DR diagnosis and grading systems, such as the Early Treatment Diabetic Retinopathy Study (ETDRS) and the International Clinical Diabetic Retinopathy (ICDR) scales, underscoring their importance for early detection and screening.[1,2] MAs are observed across both nonproliferative DR (NPDR) and proliferative DR (PDR). The presence, number, and dynamic turnover of MAs are associated with disease severity, progression, and the development of diabetic macular edema, as they represent sites of capillary wall weakening, vascular leakage, and microvascular dysfunction.[3–5] Changes in the number of MAs are also associated with treatment response, highlighting their potential role as quantitative imaging biomarkers.[6–8] Accurate detection and characterization of MAs can facilitate early diagnosis, risk stratification, and longitudinal monitoring of DR. Beyond its relevance to DR assessment, MA quantification may provide information regarding systemic microvascular status. A recent study demonstrated that chronic kidney disease was independently associated with a higher macular MA count in eyes with referable DR.[9]

Traditionally, MA assessment relies primarily on color fundus photography and fluorescein angiography.[10–13] While they are clinically useful, these modalities provide limited depth-resolved information and cannot directly capture three-dimensional (3D) MA morphology,[14–17] Optical coherence tomography (OCT) is a non-invasive 3D imaging technique that can provide high-resolution, cross-sectional images of the retina.[18] Furthermore, OCT angiography (OCTA) further offers depth-resolved flow contrast derived from repeated OCT scans at the same position.[19,20] While several publications have detected or segmented MA using *en face* OCTA,[21,22] these approaches are limited by the occlusion of some MAs, which may not be visible on *en face* OCTA. One advantage for OCT and OCTA imaging is that they are automatically co-registered. By combining these two modalities we can investigate the perfusion status in MAs. A previous study with manually delineated MAs in 3D frame by frame on cross-sectional OCT has explored the relationship between retinal fluid and MA perfusion status.[23] However, this approach required laborious manual segmentation of MAs from individual OCT B-scans by a trained grader. For MAs to a useful clinical biomarker, an accurate automated algorithm for volumetric segmentation of MAs on OCT is needed.

Despite this, few studies have explored the automated detection of MAs using OCT. Two studies attempted to identify MAs by registering fluorescein angiography with OCT images, searching for MAs within strips of cross-sectional OCT.[24,25] This approach, however, focused on the diagnosis of MAs and did not provide the depth information and segmentation of MAs on OCT.

OCT provides rich, complex data that presents challenges in processing when using traditional image processing methods. However, deep learning has demonstrated significant success in the automatic extraction of complex features from biomedical images,[26–34] including OCT data.[34–38] Despite these advancements, MA segmentation in OCT remains technically challenging because of the small size of MAs, severe class imbalance between lesions and background, and the presence of similar-appearing structures in OCT, such as large vessels and hyperreflective materials. Furthermore, MA appearance varies across retinal location, morphology, and image quality. In this study, we developed a deep learning-based framework specifically designed for volumetric MA segmentation to address these complexities by leveraging the spatial information in 3D OCT scans.

## 2. Methods

### 2.1 Data acquisition

The Institutional Review Board of Oregon Health & Science University approved this study. Informed consent was obtained from all participants, and the study adhered to the Declaration of Helsinki. Multiple repeated 3×3-mm macular OCT scans (4-10), with a transverse sampling density of 400×400 pixels, were acquired from a single eye of each participant within a continuous 5-minute session using a commercial 120-kHz spectral-domain OCT system (Solix; Visionix/Optovue, Inc., California, USA). The internal limiting membrane, nerve fiber layer, ganglion cell/inner plexiform layer, inner nuclear layer, outer plexiform layer, and the outer nuclear layer were segmented by a guided, bidirectional graph search algorithm.[35] To create motion-free volumes, multiple orthogonal scans (x-fast and y-fast) were registered.[36] Then, these volumes were registered and merged to obtain high-definition OCT volumes.[37]

### 2.2 Challenges of MA identification in OCT

On cross-sectional OCT, MAs appear as oval-shaped lesions with hyperreflective walls surrounding hyporeflective lumens (Fig. 1A, 1B, 1C, 1I).[12,23,38] Large vessels and some other lesions can resemble MAs. Large vessels have similar features to MAs on a single cross-sectional OCT (Fig. 1D), but OCT volumes can demonstrate them as tubular structures (Fig. 1G). Intraretinal hyperreflective materials, corresponding with pigment or lipid, also mimic MAs but are brighter and lack a hyporeflective lumen (Fig. 1E). Volumetrically, they appear as solid and irregular structures (Fig. 1H). Additionally, intraretinal cysts, which are discrete pockets of fluid, can appear as low-reflectance oval regions on cross-sectional OCT without hyperreflective walls (Fig. 1F). These similarities between lesions, large vessels, and MAs present significant challenges for accurate identification and segmentation of MAs in individual OCT cross-sections. To address this, we used volumetric OCT as input, which provides distinct features to reliably differentiate MAs from other structures.

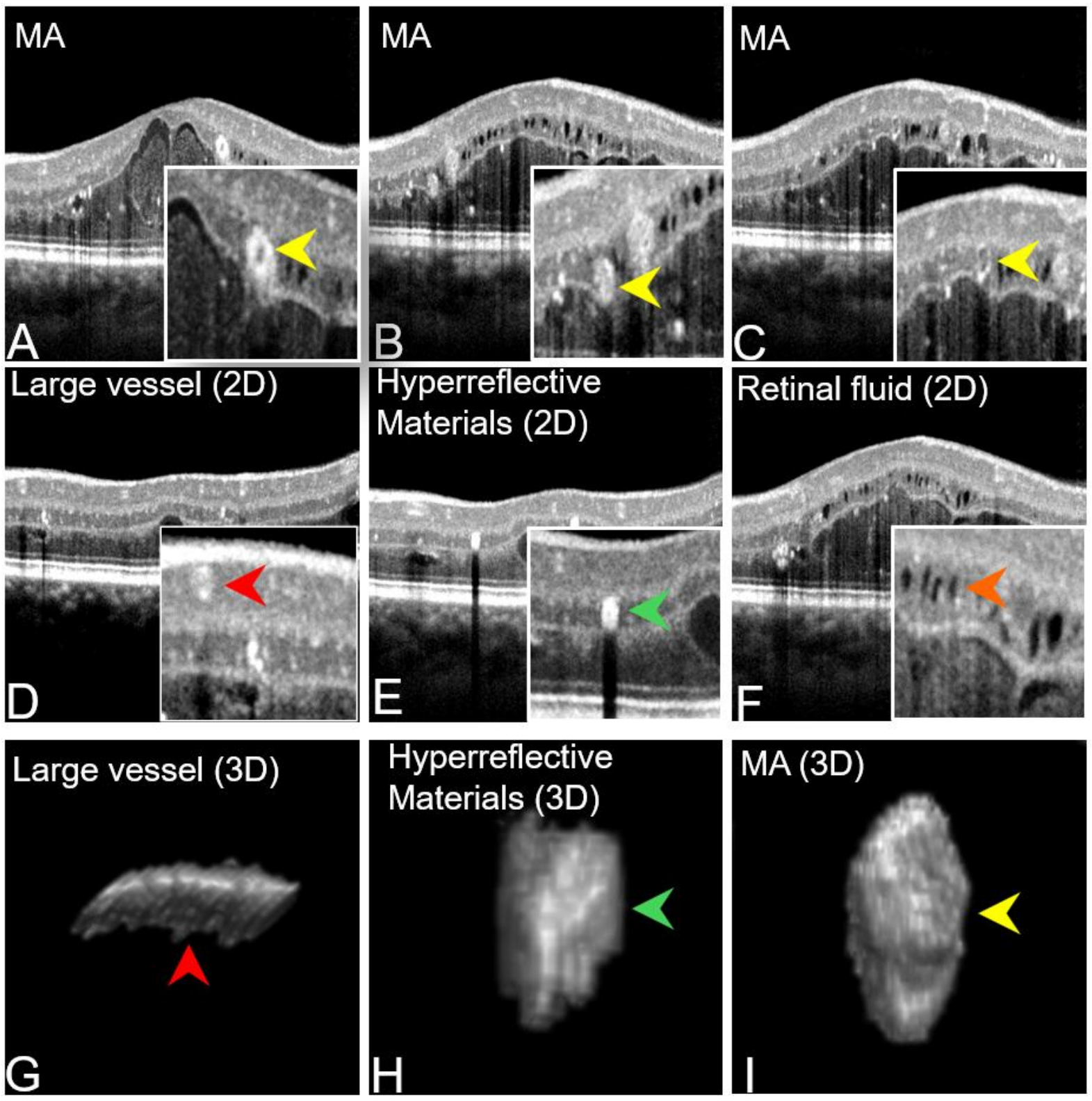

Figure 1. Microaneurysms (MAs), large vessels, hyperreflective materials, and fluid cysts on OCT. (A-C) MAs are oval-shaped lesions with hyperreflective walls and dark lumens (yellow arrows). (D) Large vessels exhibit features similar to MAs (red arrows) on cross-sectional OCT, but volumetric representation show their distinctive feature. (E) Hyperreflective materials resemble MAs but lack a dark lumen (green arrows) on OCT cross-section. (F) Fluid cysts appear as low-reflectance oval regions without hyperreflective walls (orange arrows) on OCT cross-section. (G) Large vessels appear as tubular structures within the OCT volume. (H) Hyperreflective materials manifest as solid lesions with irregular 3D shapes. (I) MAs are oval-shaped lesions in OCT volumes.

**2.3 An end-to-end three-dimensional convolutional neural network**

We designed an end-to-end 3D convolutional neural network to achieve volumetric segmentation of MAs in OCT volumes (Fig. 2). The concatenated original OCT and reflectance-inverted OCT volumes served as the input, with manually delineated 3D MA masks as the ground truth. The network is a U-Net-inspired architecture, combining an encoder-decoder structure with residual blocks. In the encoder, the network employs a series of convolutional layers followed by two types of residual blocks (standard residual modules and down-sampling residual modules).[39] These residual blocks are used to address the vanishing gradient problem and learn deeper features. During feature extraction, the spatial resolution of the OCT volumes is progressively reduced (×1, ×1/2, ×1/4), while the depth of the feature maps increases (32, 64, 96 channels). The decoder mirrors the encoder structure and incorporates up-sampling blocks to restore the spatial resolution of the feature maps. Skip connections fused feature maps from corresponding encoder layers to decoder, preserving fine structural details.

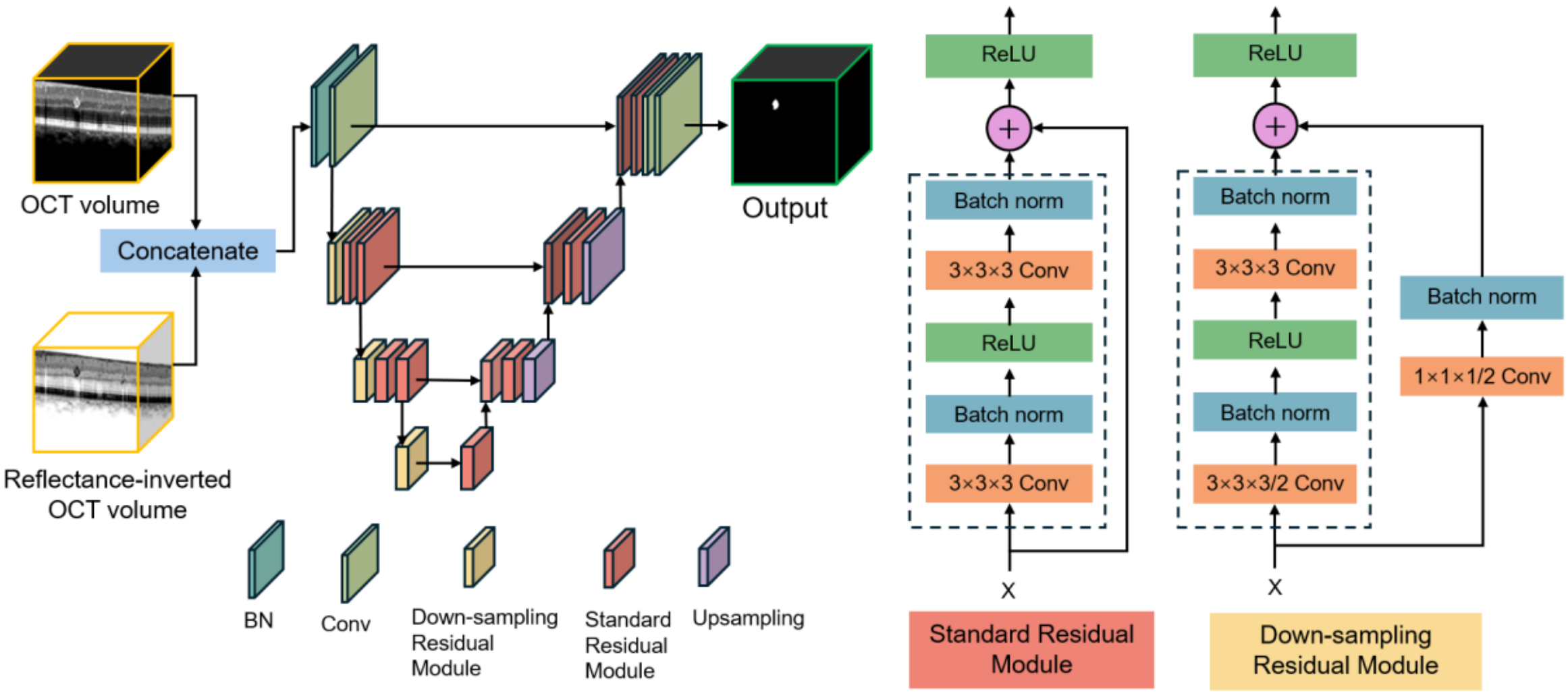

Figure 2. Schematic of the deep learning-based microaneurysm (MA) volumetric segmentation network. The input consists of the original OCT volume and its reflectance-inverted counterpart. The network follows a U-Net-like encoder-decoder architecture, with two types of residual blocks (standard residual modules and down-sampling residual modules). The network output is segmented volumetric MAs.

**2.4 Dataset**

To create the input, the OCT volume exhibiting the highest signal strength index (SSI) was selected from a series of 4–10 repeated 3×3-mm scans. This volume served as the fixed reference for registration. Subsequently, all registered volumes were averaged to produce a high-definition volume with enhanced contrast (Fig. 3A, 3B). Both the fixed single volume and the averaged volume were used as inputs to the network. For each, the original OCT volume and its reflectance-converted counterpart were concatenated and fed into the network as input (Fig. 3D, 3E). Exclusion criteria included poor OCT image quality (SSI $< 50$), severe motion or shadow artifacts, registration failure, and coexisting retinal diseases, such as branch retinal vein occlusion or central retinal vein occlusion.

To generate a provisional ground truth for training the initial model, two trained graders (J.H. and A.K.) semi-automatically delineated MAs on OCT cross-sections of averaged volumes using custom software developed by our group. A third trained grader (M.G.) reviewed these delineations to establish the human-graded provisional ground truth (Fig. 3C). Since OCT volumes were spatially registered, this provisional ground truth was directly mapped from the averaged volume to the corresponding single volume (Fig. 3F).

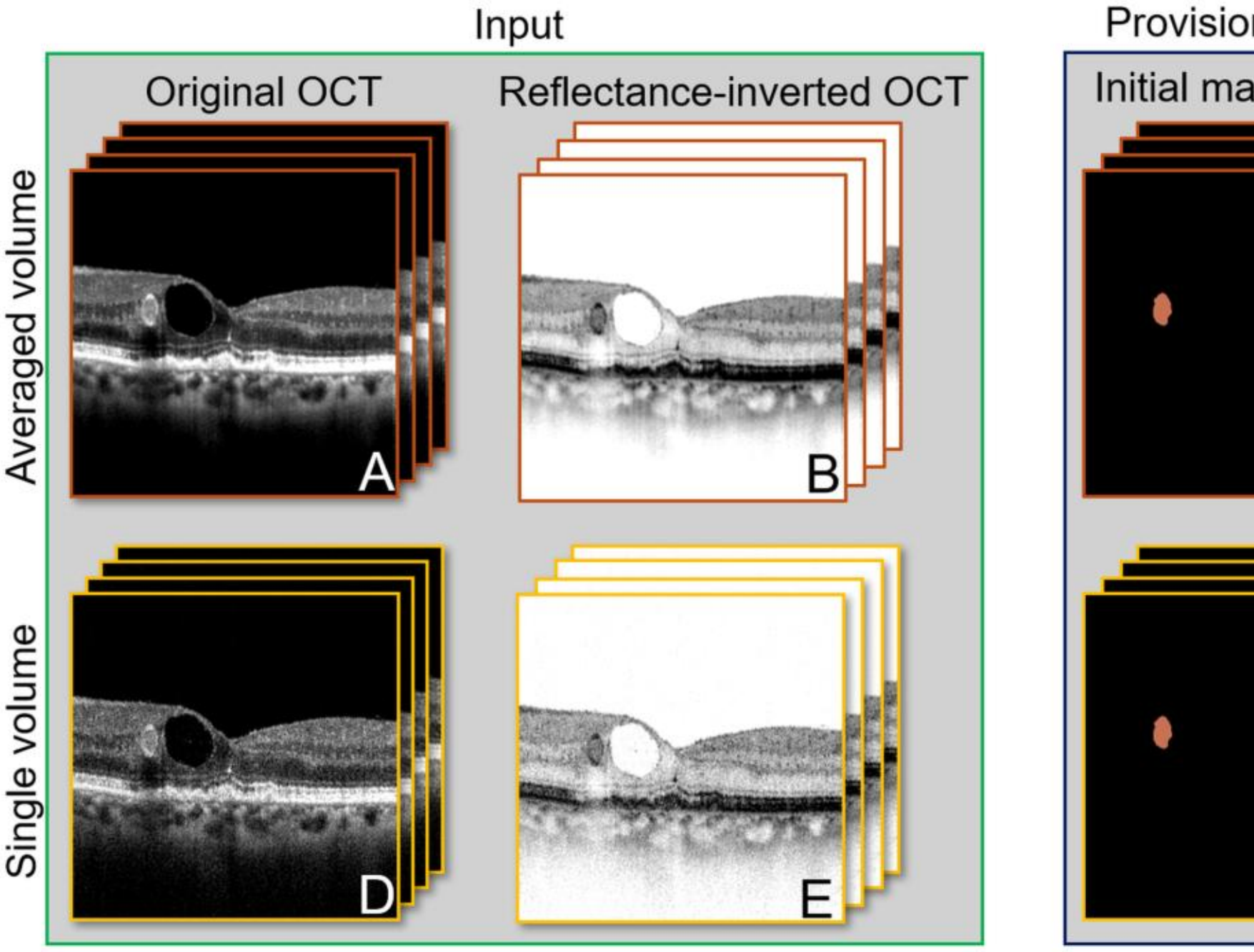

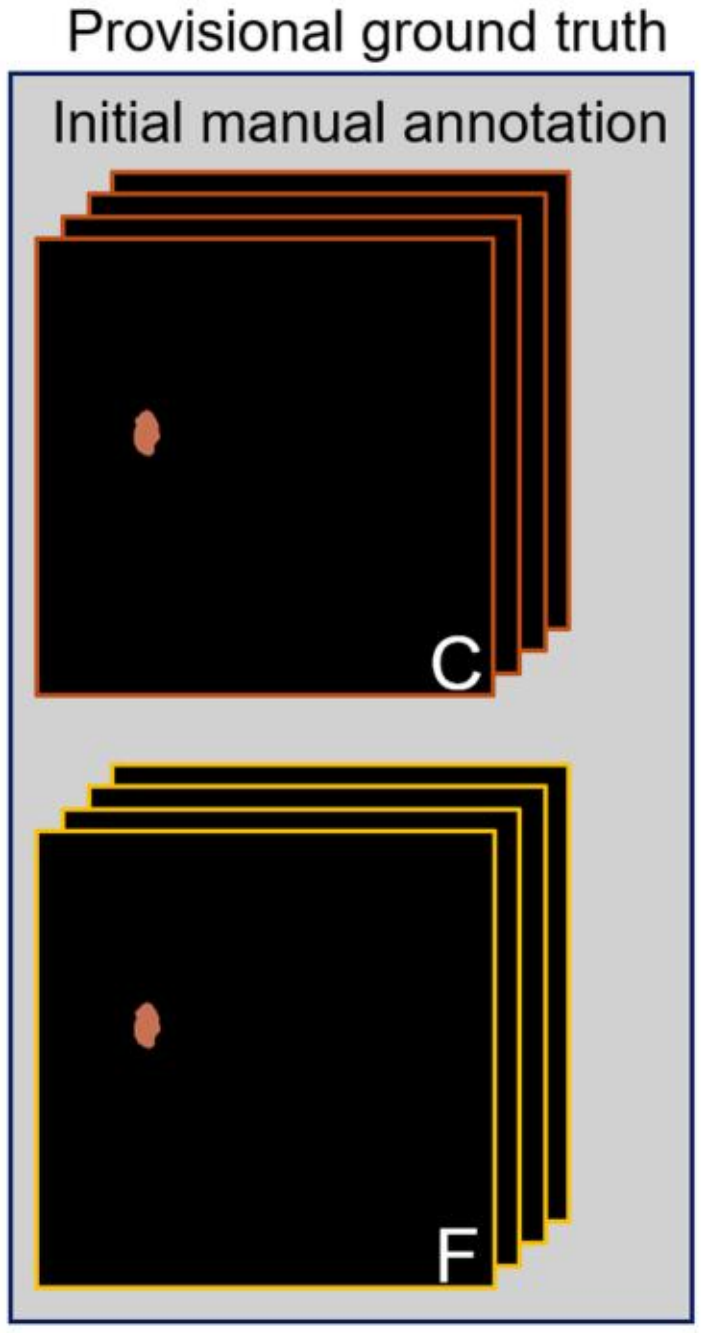

Figure 3. Preparation of inputs and provisional ground truth. (A, D) Original OCT intensity volumes for the averaged (A) and single (D) volumes. (B, E) Corresponding reflectance-inverted OCT volumes. (C) Provisional ground truth: MA masks were delineated frame-by-frame on the averaged OCT cross-sections. (F) The volumetric masks generated in (C) were applied to the registered single volumes.

To establish a robust reference standard, an initial model was trained using the provisional manual annotations, and its predictions were subsequently used as an additional quality-control tool to identify potentially missed MAs. Model-detected lesions that had been overlooked or mischaracterized by the human graders were retrospectively reviewed across multiple adjacent OCT B-scans (Fig. 4A). A candidate MA was defined as a focal round or oval lesion with a hyperreflective wall and hyporeflective lumen that could be traced across adjacent B-scans and distinguished from the continuous tubular morphology of retinal vessels, the solid appearance of hyperreflective materials, and the cavity-like configuration of intraretinal cysts. OCTA evidence of intralesional flow supported the identification of perfused MAs. For lesions without detectable OCTA flow, fundus photography and fluorescein angiography, when available, were reviewed for a corresponding small, round red lesion and focal hyperfluorescence or leakage at the same retinal location. Confirmed lesions were incorporated into the provisional annotations (Fig. 4B) to generate the final verified ground truth (Fig. 4C). Candidate lesions that remained ambiguous after volumetric and multimodal expert review were excluded. The verified test-set annotations were finalized before evaluation of the final model, and graders were masked to the final model predictions during adjudication. No test images or annotations were used for training or optimization of the final model.

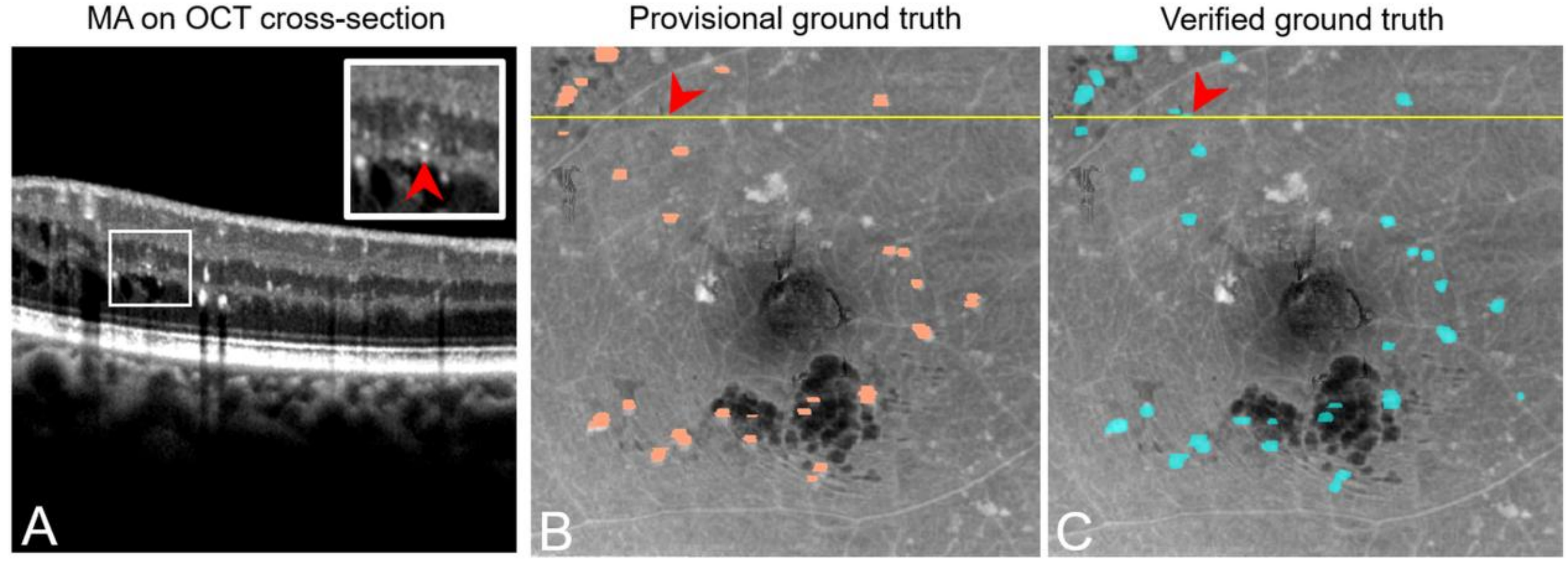

Figure 4. Creation of the verified ground truth. (A) Representative OCT cross-section showing an MA (red arrow) characterized by small size and low contrast. (B) Inner en face OCT overlaid with the provisional human-graded ground truth (orange). The red arrow indicates the location of the missed MA shown in (A). The yellow line marks the location of the corresponding cross-section. (C) Verified ground truth (cyan) overlaid on the en face OCT, demonstrating the incorporation of the previously missed MA.

The entire dataset was split into training and test sets with a ratio of 4:1. To ensure the evaluation was robust across the disease spectrum, the test set was stratified to include an equal distribution of DR severities. During training, the training set was further randomly divided into training and validation sets, also at a 4:1 ratio. To mitigate computational costs and address class imbalance in training, the original 3×3-mm volumes (640×400×400 pixels) were cropped into smaller sub-volumes of 128×128×32 pixels. Rather than random cropping, we employed a targeted sampling strategy to include challenging non-MA structures. Large vessels were automatically extracted by a previously proposed deep learning-based method that can differentiate arteries and veins.[40] Hyperreflective materials were roughly segmented using a binarization operation on OCT cross-sections. Retinal fluid was automatically segmented using another deep learning-based method that was previously reported[40].[41] The cropped sub-volumes containing large vessels, hyperreflective materials, and retinal fluid were used as negative samples, whereas the sub-volumes containing MAs served as positive samples for training the network. During testing, to address memory constraints, sub-volumes (640×400×32 pixels) encompassing the entire OCT cross-section were input to generate segmented volumetric MAs.

**2.5 Loss function**

To minimize the discrepancy between the ground truth and predicted segmentation, we used a total loss ($L_{Total}$) [Eq. (1)] that combines binary cross-entropy ($L_{BCE}$) [Eq. (2)], Dice loss ($L_{Dice}$) [Eq. (3)], and Focal Tversky loss ($L_{FT}$) [Eq. (4)].

$$L_{Total} = L_{BCE} + L_{Dice} + L_{FT} \quad [1]$$

The binary cross-entropy loss function

$$L_{BCE} = -\frac{1}{n}\sum_{i=1}^{n}(Y_i \cdot log\hat{Y}_i + (1 - Y_i) \cdot log(1 - \hat{Y}_i)) \quad [2]$$

helps measure the voxel-wise error between the predicted and actual values. Here, $Y_i$ is the ground truth of the $i^{th}$ category, and $\hat{Y}_i$ is the output prediction for the $i^{th}$ category. $L_{BCE}$ is a value between 0 and positive infinity, approaching 0 when the predicted output closely matches the ground truth. While the Dice loss

$$L_{Dice} = 1 - \frac{2TP}{2TP + FN + FP} \quad [3]$$

focuses on maximizing the overlap between the predicted segmentation and the ground truth. *TP* is true positive, *FN* is false negative, and *FP* is false positive. The Focal Tversky loss

$$L_{FT} = (1- \frac{TP + s}{TP + \alpha *FP + (1 - \alpha) *FN + s})^{r} \quad [4]$$

further enhances this by giving more weight to hard-to-segment regions. $s$ is a small smoothing constant added to both the numerator and the denominator to avoid division by zero and to ensure numerical stability. $\alpha$ is a weighting parameter that controls the relative importance of FP and FN. $r$ is the focusing parameter that adjusts the model's focus on harder-to-segment examples versus easier ones. The values for these hyperparameters were empirically determined.

### 2.6 Hyperparameters

The training batch size was set to 4. We used an Adam optimizer with an initial learning rate of 0.0001 to optimize the loss function. In the focal Tversky loss, $s, \alpha$, and $r$ were empirically set to 1.0, 0.7, and 2.0, respectively. The learning rate was reduced by a factor of 0.1 if the loss did not decrease after 5 epochs. The minimum learning rate is $10^{-8}$. Training was stopped if the loss remained unchanged for 10 epochs. Our network training was implemented using Python 3.9 with Keras (Tensorflow-backend) on a graphics processing unit (GPU) server equipped with 24G RAM and two GeForce RTX 3090 graphics cards.

### 2.7 Algorithm evaluation

We compared manual and automated identification of MA in both single and averaged volumes to assess the differences between the provisional ground truth and algorithmic grading. We evaluated the predicted results with the verified ground truths on the test dataset at multiple levels, including voxel-level MA segmentation, lesion-level MA detection, and scan-level MA diagnosis. The metrics used for evaluation included recall, precision, F1 score, and intersection over union (IoU). The Wilcoxon rank-sum test was used to determine whether these metrics differ significantly between single and averaged volumes. Metrics were compared across DR severity groups using the Kruskal–Wallis test.

We also compared the number of MAs in the different DR severities, including mild, moderate, severe NPDR and PDR. A Kolmogorov-Smirnov test was first performed to assess whether the measurements followed a normal distribution. Subsequently, a Kruskal-Wallis test was used to determine if there were differences among at least one of the groups. Dunn's post hoc tests with Bonferroni correction were then conducted to identify which specific pairs of groups showed significant differences.

### 2.8 Spatial localization of MAs

To characterize the spatial distribution of MAs, we analyzed the automatically segmented 3D MA masks across five retinal layers: the nerve fiber layer, ganglion cell/inner plexiform layer, inner nuclear layer, outer plexiform layer, and outer nuclear layer. Consistent with our previous methodology,[23] the axial locations of MAs were estimated by calculating the volumetric proportion residing within each layer. An MA was classified as present in a specific layer if more than 10% of its total volume fell within that layer's boundaries. Consequently, a single MA could be assigned to multiple adjacent layers.

### 2.9 Longitudinal analysis of MA dynamics

Two scans from visits taken six months apart were obtained from an eye with severe NPDR and macular edema. These two scans were input into our model to produce the segmented volumetric MAs. The scans were then registered to align the same region, enabling the longitudinal analysis of MA dynamics over time. OCT cross-sections were used to visualize the structural changes in MAs after six months. Retinal fluid dynamics were also analyzed over this six-month period. We also calculated the number of resolved and new MAs after treatment with anti-vascular endothelial growth factor.

## 3. Results

### 3.1 Dataset

We reviewed OCT volumes from 130 eyes of 130 patients to identify the presence of MAs. We obtained 105 OCT scans from 105 patients (age: 58.0±12.4) with DR, including 27 with mild NPDR, 30 with moderate NPDR, 30 with severe NPDR, and 18 with proliferative DR (PDR), all of which exhibited MAs (Table 1). Training and validation dataset included 85 eyes, and the test dataset included 20 eyes. In total, we cropped 2467 OCT sub-volumes from single and averaged volumes from 85 eyes with DR to train the network.

**Table 1. Distribution of the number of eyes in training and test dataset.**

| | **Training dataset** | **Test dataset** | **Total** |
|---|---|---|---|
| **Mild NPDR** | 22 | 5 | 27 |
| **Moderate NPDR** | 25 | 5 | 30 |
| **Severe NPDR** | 25 | 5 | 30 |
| **PDR** | 13 | 5 | 18 |
| **Total** | 85 | 20 | 105 |

**3.2 Automated algorithm is more sensitive to MA**

We identified four categories that could describe the difference between human and algorithm grading on single and averaged volumes, as follows: (1) Both our algorithm and human graders identified MAs on single and averaged cross-sectional OCT scans. (2) Both identified MAs on the averaged volumes (Fig. 5B) but the algorithm did not detect them on single volumes due to low resolution and contrast (Fig. 5A). (3) Our algorithm detected MAs on averaged volumes but not on single volumes, whereas human graders did not identify MAs on averaged volumes due to their small size (Fig. 5C, 5D). (4) The algorithm detected MAs on both single and averaged volumes, while human graders failed to detect them (Fig. 5E, 5F). In 20 test eyes, there were 13 (5.3%) of 247 MAs missed by human graders on averaged volumes, while the initial model trained on provisional ground truth was able to detect them on averaged volumes.

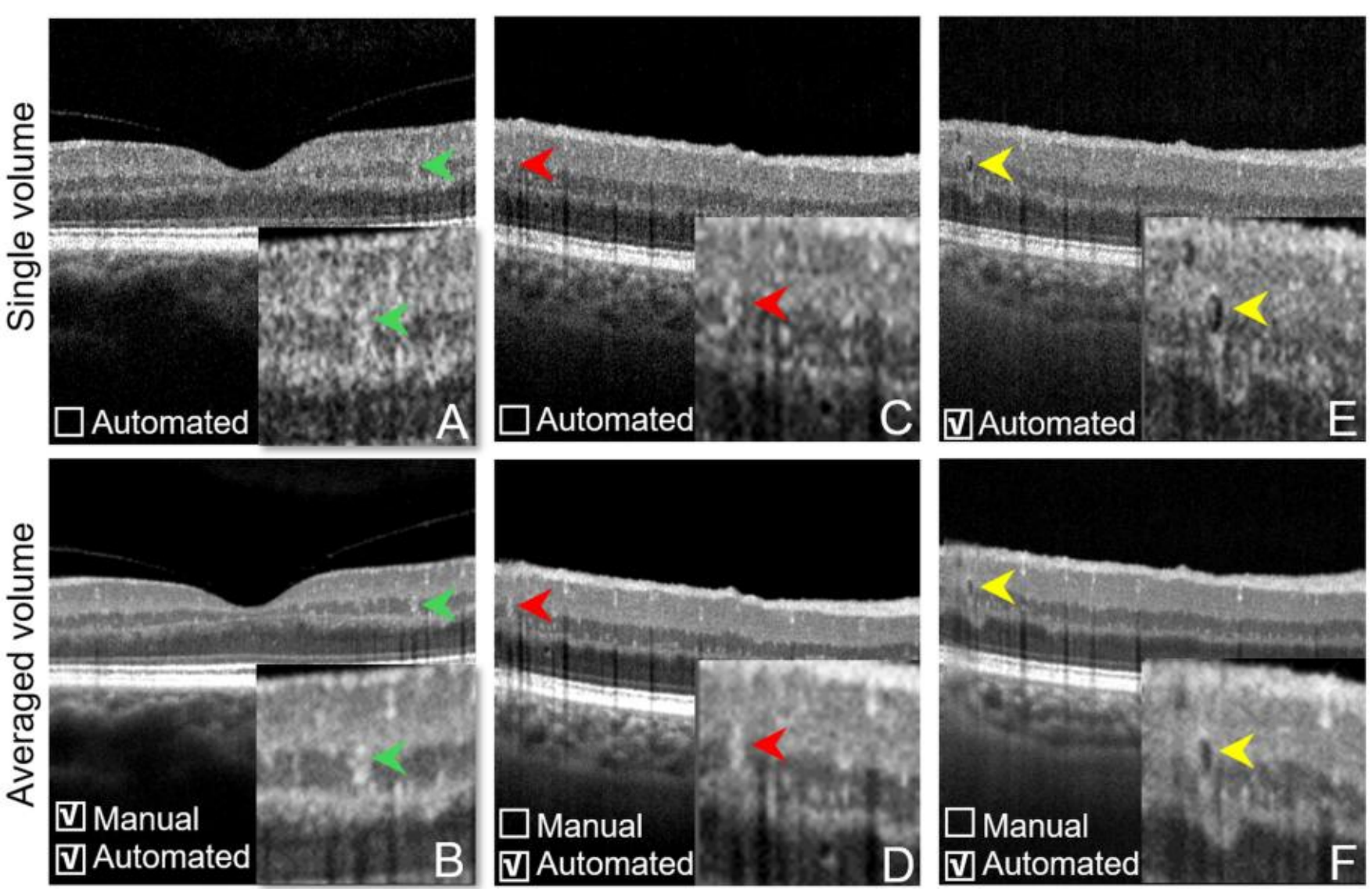

Figure 5. Comparison of MA detection performance between the automated algorithm and provisional human ground truth. (A-B) A small, low-contrast MA (green arrow) on the single volume (A) was missed by the algorithm. However, averaging enhanced the contrast (B), allowing identification by both the algorithm and human graders. (C–D) A small MA (red arrow) missed by the algorithm on the single volume (C) was detected by the algorithm on the averaged volume (D) but missed by human graders. (E–

F) An MA (yellow arrow) with indistinct boundary on the single volume (E) was successfully detected by the algorithm on both single and averaged (F) volumes, despite being missed by human graders.

### 3.3 Voxel-level segmentation accuracy

The segmentation model demonstrated robust voxel-level performance across both single and averaged volumetric datasets (Table 2, Fig. 6). Quantitative analysis revealed that averaged volumes achieved significantly higher accuracy than single volumes across all evaluated metrics ($P < 0.05$). Specifically, the use of averaged scans yielded an improvement of approximately 10% in the IoU and an 8.5% increase in the recall rate compared to single-scan inputs.

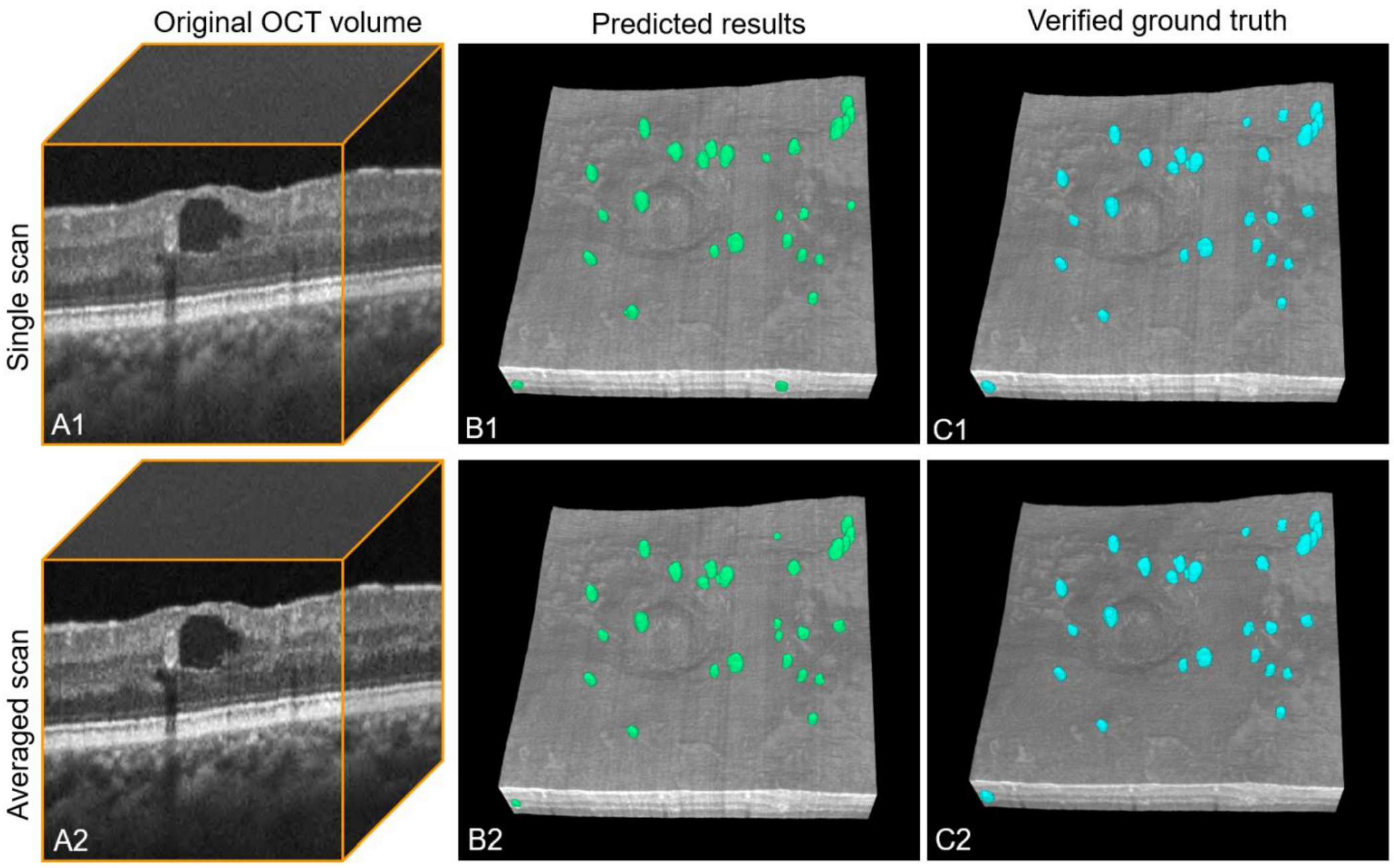

Figure 6. Volumetric segmentation microaneurysms (MAs) on single and averaged OCT scans. (A1, A2) Original OCT volume of single and averaged volumes, respectively. (B1, B2) Automated predicted results showing segmented MAs (green) on single and averaged volumes. (C1, C2) Verified ground truth (cyan) refined based on the network's output to ensure gold-standard accuracy.

**Table 2. Voxel-level performance of MA segmentation on the test dataset (N = 20).**

| MA segmentation | Recall (%) | Precision (%) | F1 score (%) | Intersection over Union (%) |
|---|---|---|---|---|
| **Single volumes** | 78.0±11.7 | 82.2±11.6 | 79.2±9.3 | 66.4±12.2 |
| **Averaged volumes** | 86.5±12.0 | 87.1±9.3 | 86.1±8.3 | 76.6±13.3 |

These metrics showed significant differences between single and averaged volumes ($P<0.05$).

### 3.4 Lesion-level and scan-level accuracy

The algorithm demonstrated robust lesion-level performance across both single and averaged volumes. As summarized in Table 3, the model consistently achieved performance metrics exceeding 90% when evaluated against verified ground truth (Table 3). Notably, while averaged volumes exhibited marginally higher recall, F1 score, and IoU alongside slightly lower precision compared to single volumes, these performance variations were not statistically significant ($P > 0.05$).

At the scan level, the algorithm demonstrated exceptional diagnostic reliability, achieving 97.5% accuracy in detecting the presence of MAs across the test set of 20 healthy eyes and 20 eyes with DR. This high level of detection accuracy remained consistent regardless of whether single or averaged volume inputs were utilized.

Model performance at different evaluation levels did not differ significantly across DR severity groups, including mild, moderate, and severe NPDR and PDR, in any pairwise comparison within the test dataset (all $P > 0.05$).

**Table 3. Lesion-level performance of MA detection on the test dataset (N = 20).**

| MA detection | Recall (%) | Precision (%) | F1 score (%) | Intersection over Union (%) |
|---|---|---|---|---|
| **Single volume** | 99.3±2.3 | 96.6±7.9 | 96.0±8.1 | 93.4±2.0 |
| **Averaged volume** | 99.5±2.0 | 95.4±10.1 | 97.1±6.2 | 95.0±10.4 |

These metrics showed no significant differences between single and averaged volumes (P>0.05).

### 3.5 MA comparison in different DR severities

MA counts were calculated based on the verified ground truth and compared across the entire study population, which included eyes with mild NPDR (n=27), moderate NPDR (n=30), severe NPDR (n=30), and PDR (n=18). A significant difference was observed in the number of MAs between mild and moderate NPDR (7.4±1.5 vs. 12.7±1.6, $P = 0.014$). However, no significant differences were found between the other pairwise groups ($P > 0.05$) (Fig.7).

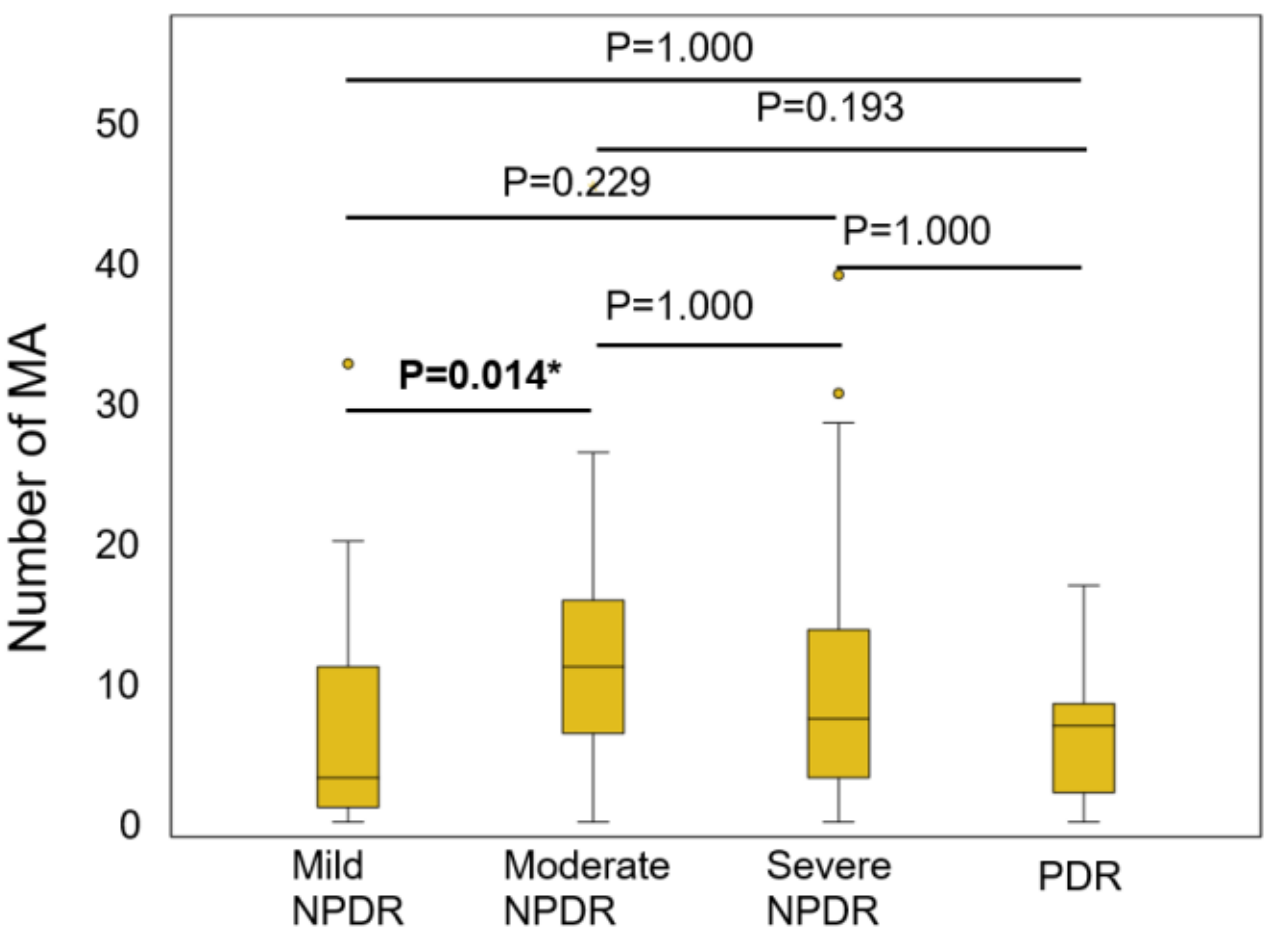

Figure 7. Comparison of MA numbers in different DR severities. There was a significant difference between mild and moderate NPDR.

### 3.6 Spatial localization and visualization of MAs

Consistent with our prior manual-grading findings,[23] automatically detected MAs were predominantly located in the inner nuclear layer and outer plexiform layer; these two layers contained the highest MA volume (Fig. 8A). To further illustrate this distribution, MA location within the retina was visualized using a depth-resolved colormap (Fig. 8B).

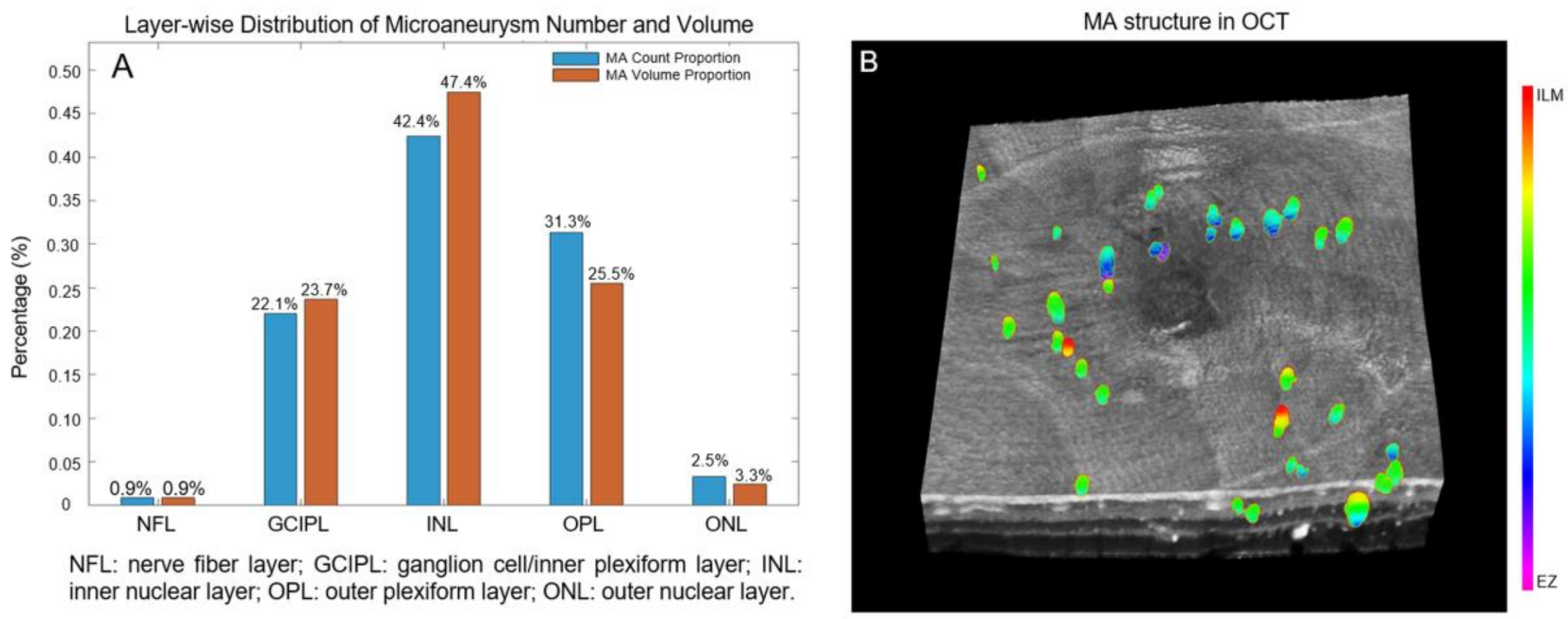

Figure 8. Layer-specific analysis of microaneurysms (MAs) number and volume and visualization of MA structure using OCT. (A) Distribution of MA count and volume across retinal layers. Inner nuclear layer and outer plexiform layer show the highest proportion. (B) Three-dimensional visualization of MA structure derived from OCT, rendered with a depth-encoded colormap to indicate axial location.

### 3.7 Longitudinal analysis of MA dynamics (case illustration)

We can observe the dynamic changes in MAs through follow-up scans. In one case of severe NPDR with macular edema, cross-sectional OCT revealed a clear MA at baseline (Fig. 9A1), which was absent after 6 injections of intravitreal aflibercept (Eylea) anti-VEGF agents over six months (Fig. 9A2). Both fluid volume and the overall number of MAs significantly decreased following treatment, as shown in volumetric renderings (Fig. 9B1, B2). In this eye, 11 MAs disappeared after 6 months of treatment (Fig. 9C1), 3 new MAs appeared (Fig. 9C2), and 26 persistent MAs presented at both baseline and follow-up.

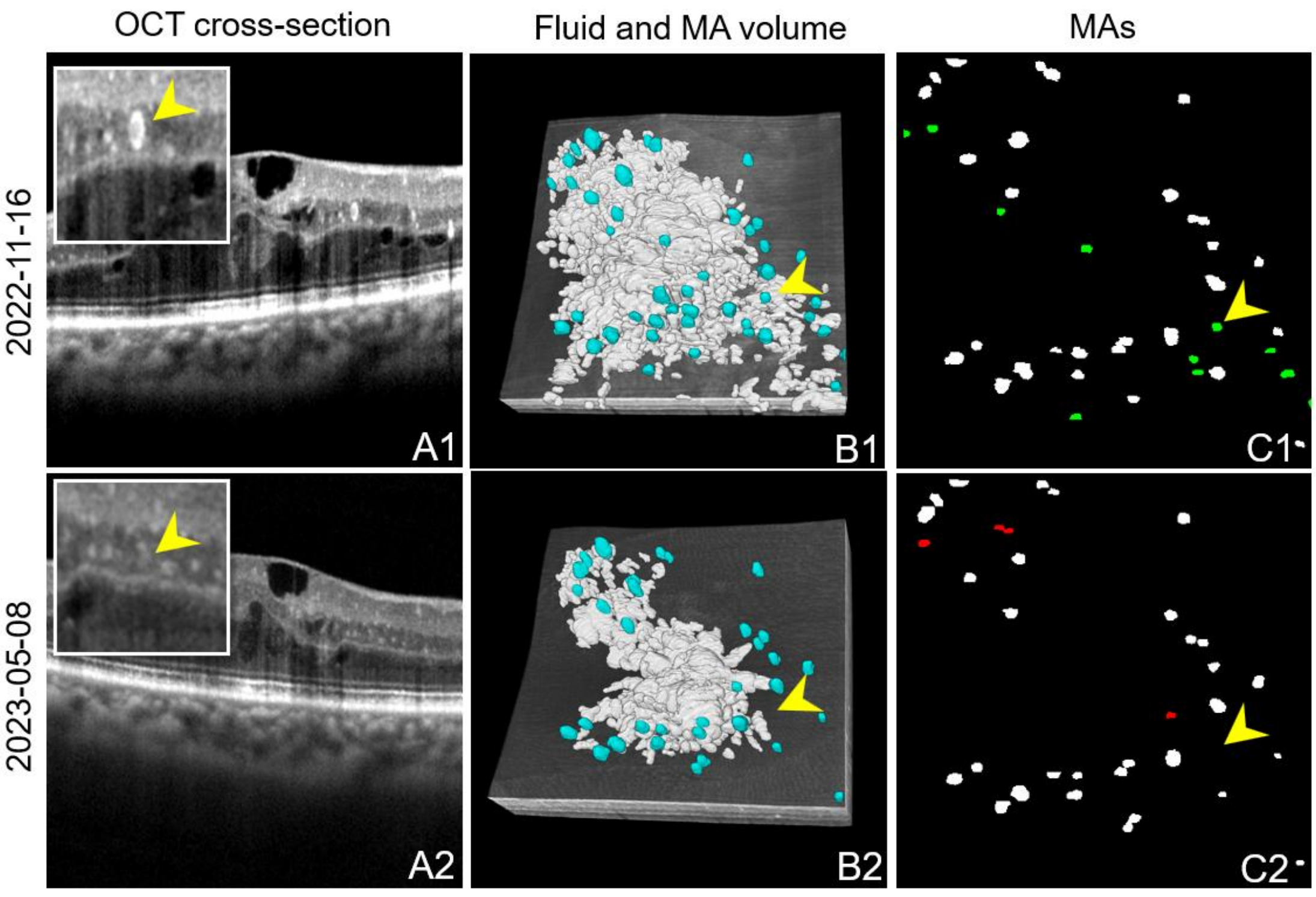

Figure 9. Dynamic changes in microaneurysms (MAs) and fluid volume in a case of severe NPDR with macular edema over 6 months of treatment. (A1, A2) Cross-sectional OCT images show an MA structure (yellow arrow) at baseline (A1) that is no longer present after 6 months of treatment (A2). (B1, B2) Three-dimensional volumetric renderings show a significant reduction in fluid volume (white) and MA (cyan) number after 6 months of treatment. (C1, C2) Eleven MAs (green) resolved after treatment (C1), three new MAs (red) appeared at follow-up (C2), and 26 persistent MAs (white) present at both timepoints.

## 4. Discussion

Microaneurysms are a key feature of DR and have been shown to predict disease progression. However, their identification and segmentation on OCT is challenging due to their small size. Advances in high-resolution OCT have revealed distinct MA features, enabling detection through advanced deep learning methods. In this study, we developed an end-to-end 3D deep learning network for automated volumetric segmentation of MAs on OCT with a high accuracy, potentially enabling clinical use of MA count as a practical biomarker. By averaging multiple high-resolution 3×3-mm scans, we generated high-definition OCT volumes with enhanced contrast and reduced noise, improving MA identification and precise ground truth delineation on cross-sectional OCT. To the best of our knowledge, this is the first end-to-end automated volumetric segmentation algorithm for identification of MAs on OCT.

Our network achieved robust performance in MA segmentation by leveraging three synergistic strategies to resolve morphological ambiguities. First, we utilized 3D OCT volumes rather than 2D cross-sections; unlike 2D images, where MAs are easily confused with large vessels, retinal fluid, and hyperreflective materials, volumetric data provides the distinct features for accurate differentiation. Second, to address class imbalance and further suppress false positives, we implemented a targeted hard-negative sampling strategy. By enriching the training data with cropped regions of these specific confounding structures, we compelled the network to learn highly discriminative features. Third, the concatenated original OCT and reflectance-inverted OCT volumes that served as input can extract rich and complementary feature representations. Although both inputs originate from the same anatomy, they exhibit different intensity distributions. Some MAs appear faint or exhibit low contrast in the original OCT, but their visibility is enhanced after reflectance inversion. This combination of volumetric context, adversarial-style training data sampling, and complementary feature representations significantly enhanced the model's robustness and generalizability.

The comparative analysis of the results presented in Tables 2 and 3 reveals a distinct performance gap between voxel-level precision and lesion-level detection reliability. At the voxel level, volume averaging yields significant improvements across all metrics, a nearly 6.9% increase in F1 score (from 79.2% to 86.1%) and a 10% gain in IoU (from 66.4% to 76.6%). This suggests that the improved signal-to-noise ratio in averaged volumes enhanced the definition of the precise boundaries of MAs, particularly those with low contrast that are easily obscured by noise in single scans. However, at the lesion level, the impact of averaging is far less pronounced. While the algorithm achieves high recall on both single and averaged datasets, the differences in precision and F1 score are not statistically significant. This indicates that while averaging helps the segmentation mask and refines the morphological shape of the lesion, the model is capable of detecting the presence of MAs from single-volume inputs, that tend to have lower signal to noise ratio.

Although averaged volumes significantly improve voxel-level segmentation, repeated acquisition of 4-10 scans and the subsequent registration processing substantially increased imaging and processing time. In our protocol, acquiring 4–10 repeated 3×3-mm scans for a single eye required approximately 2-6 minutes of continuous imaging. While this approach is valuable for generating validated ground truth and detailed morphologic analysis, it is less practical for busy clinical care. Because our model achieves comparable scan-level diagnostic accuracy with single and averaged volumes (97.5%), single-volume imaging may offer a more efficient and clinically practical approach for routine DR screening and monitoring.

A significant difference in MA counts was observed between mild and moderate NPDR; however, no significant differences were identified in other pairwise comparisons. This observation may be attributed to the progression of capillary non-perfusion and ischemia in severe NPDR and PDR, which may lead to a reduction of MAs.[42,43] One previous publication also demonstrated that disappearance of MAs mainly occurred in individuals that developed clinically significant macular edema and PDR outcomes[44]. Volumetric and numerical analyses of the detected MAs are highly consistent with our previous findings.[23] Specifically, MAs were predominantly localized within the inner nuclear layer and outer plexiform layer. This consistency further validates that our algorithm is reliable and robust for MA segmentation. Using our automated segmentation method, we also conducted a longitudinal analysis of MA dynamics and intraretinal fluid volume in an eye with severe NPDR and macular edema under treatment. After 6 months of treatment, both MA count and fluid volume are greatly reduced. Although this illustrative case demonstrates the feasibility of longitudinal lesion-level analysis, it is insufficient to establish clinical utility or treatment efficacy. Therefore, the proposed method should currently be considered a research tool for automated 3D quantification and longitudinal characterization of MAs. Larger longitudinal studies are needed to determine whether it provides incremental value beyond existing multimodal assessment or can inform clinical decision-making.

The utility of MA quantification may extend beyond conventional DR grading. Objective measurements of MA burden, morphology, perfusion, and turnover may better characterize microvascular instability and longitudinal disease activity than lesion presence alone. Furthermore, because retinal MAs may reflect systemic microvascular injury, their quantitative assessment could provide insights into both ocular and systemic conditions, such as chronic kidney disease. Further longitudinal studies are needed to determine their prognostic and clinical value.

A potential limitation is that model-assisted refinement of the ground truth may have introduced confirmation bias by preferentially identifying MAs with features learned from the provisional annotations, potentially overrepresenting model-recognizable lesions while missing lesions overlooked by both graders and the model. In addition, the algorithm was developed and validated using data from a single center and a single OCT device, which may limit its generalizability. Future studies should include validation using external, expert-annotated, multicenter, and multi-device datasets.

In conclusion, an automated deep learning method was developed to volumetrically segment MAs on OCT, achieving strong performance and enabling reliable quantification and characterization of MAs. The method may serve as a research tool for studying MA morphology and longitudinal dynamics, potentially providing clinically relevant insights into DR progression and treatment response.